\documentstyle[epsf]{mn}
\newif\ifAMStwofonts

\def\simlt{\lower.5ex\hbox{$\; \buildrel < \over \sim \;$}}
\def\simgt{\lower.5ex\hbox{$\; \buildrel > \over \sim \;$}}

\def\be{\begin{equation}}
\def\ee{\end{equation}}

\def\bee{\begin{eqnarray}}
\def\eee{\end{eqnarray}}

  \ifCUPmtlplainloaded \else
  \fi
  \ifAMStwofonts
    \ifCUPmtlplainloaded \else
      \NewSymbolFont{upmath} {eurm10}
      \NewSymbolFont{AMSa} {msam10}
      \NewMathSymbol{\upi}     {0}{upmath}{19}
      \NewMathSymbol{\umu}     {0}{upmath}{16}
      \NewMathSymbol{\upartial}{0}{upmath}{40}
      \NewMathSymbol{\leqslant}{3}{AMSa}{36}
      \NewMathSymbol{\geqslant}{3}{AMSa}{3E}

    \fi
  \fi
  \addtoversion{normal}{\mathit}{cmr}{m}{it}
  \addtoversion{bold}{\mathit}{cmr}{bx}{it}

  \ifAMStwofonts
    \ifCUPmtlplainloaded \else
      \UseAMStwoboldmath
      \makeatletter
      \new@mathgroup\upmath@group
      \define@mathgroup\mv@normal\upmath@group{eur}{m}{n}
      \define@mathgroup\mv@bold\upmath@group{eur}{b}{n}
      \edef\UPM{\hexnumber\upmath@group}
      \new@mathgroup\amsa@group
      \define@mathgroup\mv@normal\amsa@group{msa}{m}{n}
      \define@mathgroup\mv@bold\amsa@group{msa}{m}{n}
      \edef\AMSa{\hexnumber\amsa@group}
      \makeatother
      \mathchardef\upi="0\UPM19
      \mathchardef\umu="0\UPM16
      \mathchardef\upartial="0\UPM40
      \mathchardef\leqslant="3\AMSa36
      \mathchardef\geqslant="3\AMSa3E

    \fi
  \fi

  \newcommand{\lsim}{\stackrel{<}{\sim}}

  \ifAMStwofonts
    \ifCUPmtlplainloaded \else
      \DeclareSymbolFont{UPM}{U}{eur}{m}{n}
      \SetSymbolFont{UPM}{bold}{U}{eur}{b}{n}
      \DeclareSymbolFont{AMSa}{U}{msa}{m}{n}
      \DeclareMathSymbol{\upi}{0}{UPM}{"19}
      \DeclareMathSymbol{\umu}{0}{UPM}{"16}
      \DeclareMathSymbol{\upartial}{0}{UPM}{"40}
      \DeclareMathSymbol{\leqslant}{3}{AMSa}{"36}
      \DeclareMathSymbol{\geqslant}{3}{AMSa}{"3E}

    \fi
  \fi
\ifCUPmtlplainloaded \else
  \ifAMStwofonts \else
    \def\upi{\pi}
    \def\umu{\mu}
    \def\upartial{\partial}
  \fi
\fi

\title[Formation of UCDs and dE(N)s]
{The formation of ultra-compact dwarf galaxies and nucleated dwarf galaxies}
\author[Tobias Goerdt et al.] {\parbox[t]{\textwidth}{Tobias
Goerdt$^{1,2}$\thanks{tgoerdt@phys.huji.ac.il}, Ben Moore$^1$, Stelios
Kazantzidis$^3$,\\ Tobias Kaufmann$^4$, Andrea V. Macci\`o$^5$ and Joachim
Stadel$^1$} \\ \vspace*{3pt} \\
$^1$Institut f\"ur Theoretische Physik, Universit\"at Z\"urich,
Winterthurerstrasse 190, CH-8057 Z\"urich, Schweiz \\ 
$^2$Racah Institute of Physics, The Hebrew University, Jerusalem 91904, Israel
\\
$^3$Kavli Institute for Particle Astrophysics and Cosmology, Stanford
University, 2575 Sand Hill Rd, Menlo Park, CA 94025 USA\\
$^4$Center for Cosmology, Department of Physics and Astronomy, University of
California, Irvine, CA 92697, USA \\
$^5$Max-Planck-Institut f\"ur Astronomie, K\"onigstuhl 17, D-69117 Heidelberg,
Deutschland\\
}

\date{Draft version \today}
\pagerange{\pageref{firstpage}--\pageref{lastpage}}
\pubyear{2007}

\begin{document}

\maketitle

\label{firstpage}

\begin{abstract}
Ultra compact dwarf galaxies (UCDs) have similar properties as massive 
globular clusters or the nuclei of nucleated galaxies. Recent observations 
suggesting a high dark matter content and a steep spatial distribution within
groups and clusters provide new clues as to their origins. We perform
high-resolution $N$-body / smoothed particle hydrodynamics simulations designed
to elucidate two possible formation mechanisms for these systems: the merging
of globular clusters in the centre of a dark matter halo, or the massively
stripped remnant of a nucleated galaxy. Both models produce density profiles as
well as the half light radii that can fit the observational constraints.
However, we show that the first scenario results to UCDs that are underluminous
and contain no dark matter. This is because the sinking process ejects most of
the dark matter particles from the halo centre. Stripped nuclei give a more
promising explanation, especially if the nuclei form via the sinking of gas,
funneled down inner galactic bars, since this process enhances the central dark
matter content. Even when the entire disk is tidally stripped away, the nucleus
stays intact and can remain dark matter dominated even after severe stripping.
Total galaxy disruption beyond the nuclei only occurs on certain orbits and
depends on the amount of dissipation during nuclei formation. By comparing the
total disruption of CDM subhaloes in a cluster potential we demonstrate that
this model also leads to the observed spatial distribution of UCDs which can be
tested in more detail with larger data sets.
\end{abstract}


\section{Introduction}
A new population of subluminous and extremely compact objects have been
recently discovered in cluster and group environments (Drinkwater et al. 2000;
Phillips et al. 2001) These ultra compact dwarf galaxies (hereafter UCDs) 
are dynamically distinct systems having intrinsic sizes $\lsim$ 100 pc and 
absolute magnitudes in the $B$-band in the range from $-13$ to $-11$ 
placing them in the lower range of dwarf galaxy luminosities (Mateo 1998). 
A number of different scenarios have been proposed for origin of UCDs 
(e.g. Evstigneeva et al. 2007 and references within): (a)\,They are simply 
very big and luminous globular clusters, (b)\,they are nucleated dwarf 
galaxies, (c)\,they are the resulting objects of the coalescence of 
several globular clusters, (d)\,they are the remnants of stripped disk 
galaxies. In this work we compare the last two scenarios.

Oh and Lin (2000), Oh, Lin and Richer (2000) as well as Fellhauer and Kroupa
(2002) investigated the third formation scenario. The latter authors performed
$N$-body simulations of the merging of several globular clusters and argued
that the resulting object resembled a UCD. Their simulations initiated with a
supercluster, an accumulation of up to fifty globular clusters, orbiting inside
a host galaxy on fairly elliptical orbits and with apo-galactic distances
around 20\,kpc. The globular clusters merged within the supercluster giving
rise to an object with values for surface brightness and absolute bolometric
luminosity comparable to UCDs.

The last formation scenario has been discussed in Oh, Lin \& Aarseth (1995) as
well as Bekki et al. (2001, 2003). The latter authors performed numerical
simulations of the dynamical evolution of nucleated dwarf galaxies orbiting
inside NGC 1399 and the Fornax cluster. Adopting a plausible scaling relation
for dwarf galaxies, Bekki et al. found that the outer stellar envelope of a
nucleated dwarf was totally removed by tidally stripping over the course of
several passages from the central region of their host. The nucleus is so dense
that it will always survive the tidal field of a group or cluster potential. By
construction in the initial conditions, the size and luminosity of the remnant
were similar to those observed for UCDs and the host galaxy was a Plummer model
halo which has a constant density core and therefore easy to tidally disrupt
leaving behind the central nucleus which would be associated with a UCD galaxy.

The main goal of the present study is to shed light into the formation of UCDs
investigating the last two formation scenarios in more detail and with more
realistic initial conditions. First, we adopt the Fellhauer and Kroupa (2002)
model and combine it with the cold dark matter model (CDM) paradigm. We assume
that globular clusters form and subsequently orbit around dark matter haloes
having masses comparable to that of the Fornax dwarf spheroidal and containing
no other baryonic matter. Due to dynamical friction, these globular cluster
would spiral in towards the centre of the halo. We perform collisionless
$N$-body simulations of this process and show that the object resulting from
the coalescence of the globular clusters at the halo centre resembles a UCD.
As we demonstrate below, the above model suffers from two major drawbacks
which rule out its applicability for UCD formation.

The first issue is related to the observed high M/L ratio (between 6 and 9 in
solar units) recently reported for UCDs \cite{ACS}. One has to note here that
the M/L ratio of UCDs is still debated (Micheal Drinkwater, private
communication, compare Evstigneeva et al. 2007 who quote a M/L for UCDs 3 and 5
in solar units). Such high values can probably not be achieved by the above
mechanism. This is because sinking globular clusters will expel most of the
dark matter particles from the halo centre \cite{amr,merritt,goerdt}. If there
is no dark matter in UCDs, Fellhauer and Kroupa (2006) describe a possible way
to enhance the mass-to-light ratios of UCDs through tidal interactions. The
second problem with this scenario is related to the total luminosity of an UCD.
At least today, dark matter halos with a sufficient number of globular clusters
to produce such bright objects are very rare
\cite{sharina,boeker1}.

Second, we examine the scenario of Bekki et al. (2001, 2003) which is based on
the hypothesis that UCDs are remnants of stripped nucleated galaxies. This
mechanism has also been discussed by Kazantzidis et al. (2003). We test this
model using SPH simulations of a low mass galaxy which forms a nucleus via gas
inflow to the inner $\approx 100$ parsecs. Once this galaxy is placed on a
critical disrupting orbit within a cluster potential, the surviving nucleus is
in excellent agreement with the latest observational constraints for UCDs,
including their dark matter content and spatial distribution.

The paper is organised as follows. In section 2 we describe the globular
cluster numerical simulations and compare the properties of the resulting
object with those of UCDs. Section 3 contains results from the simulations
of tidal stripping of disk galaxies inside a host cluster potential.
Finally, in Section 4 we summarise our main conclusions.

\section{The Globular Cluster Simulations}
The globular cluster simulations were performed with \textsc{Pkdgrav2} a
multi stepping, parallel $N$-body tree code \cite{stadel}. We create an NFW
\cite{nfw} halo, employing the technique developed by Kazantzidis, Magorrian \&
Moore (2004), which has the following density profile:
\begin{equation}
\rho(r)=\frac{\rho_0}{r / r_{\rm s}\left[1 + \left(r / r_{\rm s} \right)
\right]^2}.
\label{eqnfw}
\end{equation}
In our case $\rho_0 = 242$\,M$_\odot$/pc$^3$ and $r_{\rm s} = 1.5$\,kpc. The
halo has a virial mass of $1.5 \times 10^9$\,M$_{\odot}$. The concentration
parameter is 20 which is the typical value for halos of this initial mass. To
increase mass resolution in the region of interest, we divide the halo into
three shells \cite{zemp} each of which contains $10^5$ particles. The
innermost shell has 100\,pc radius. The second shell is between 100 and
500\,pc while the third shell contains the rest of the halo. The softening
lengths for these shells are 1, 10 and 100\,pc respectively. This shell model
allows us to resolve the detailed kinematics within the central few parsecs
whilst retaining the global structure of the halo out to its virial radius of
29.39 kpc.

We use ten globular clusters consisting of $10^5$ particles and being
represented by the King model \cite{king,michie,bodenheimer}
\begin{eqnarray}
\rho(\Psi) &=& \rho_1 \exp \left({\Psi \over \sigma^2} \right) {\rm erf}
\left({\sqrt{\Psi} \over \sigma} \right) \nonumber \\ &-&  \rho_1 \sqrt{4\Psi
\over \pi \sigma^2} \left(1+{2 \Psi \over 3 \sigma^2} \right).
\end{eqnarray}
Each globular cluster is constructed with a W$_0 = \Psi(0) / \sigma^2$
parameter of 6, a total mass of $4.2 \times 10^5$\,M$_{\odot}$, a central
velocity dispersion of 11\,km/s, and an absolute magnitudes of -8.5, assuming a
mass to light ratio of 2 \cite{boeker2,walcher}. We use 1\,pc for its
gravitational softening length. The ten globular clusters are initially
distributed within the halo between 20\,pc and 1000\,pc. They are spatially
distributed according to $\rho (r) \propto r^{-4}$, which is in agreement with
observations of globular cluster populations in dwarf galaxies \cite{sharina}.
We randomly place the globular clusters on circular orbits (We also tried
different orbital and spatial distributions including $r^{-3}$ and $r^{0}$ with
similar results as in the standard $r^{-4}$ case.).

The dwarf halo which contained the globular clusters was placed on a static NFW
potential which corresponds to a cluster halo with virial mass of $M_{\rm vir}
= 10^{14}$\,M$_{\odot}$ and concentration of $c = 6$. The halo was placed
on an initial distance of 100\,kpc from the centre of the potential and was
given a velocity of 650\,km/s, respectively. These values result to an
eccentric orbit with an apocentre of 500\,kpc. For a second simulation we put
fifty of these globular cluster in the aforementioned halo and let them spiral
into the centre. This second simulation identical to the first one, just with
fifty instead of ten globular clusters.

All ten globular clusters in our first simulation merge to a single object
within 0.4\,Gyr. The associated timescale of this evolution can be calculated
using the Chandrasekhar dynamical friction formula \cite{chandrasekhar}.
\begin{eqnarray}
{{\rm d}r \over {\rm d}t} & = & -{4 \pi {\rm ln} \Lambda (r) \rho(r) G^2
M_{\rm GC} r \over v^2_{\rm c}(r) {{\rm d}\left[rv_{\rm c}(r)\right] /
{\rm d}r}} \left\{{\rm erf}\left[{v_{\rm c}(r) \over \sqrt{2} \sigma(r)}\right]
\right. \nonumber \\ & - & \left.{2 v_{\rm c}(r) \over \sqrt{2 \pi} \sigma(r)}
{\rm exp}\left[{-v_{\rm c}^2(r) \over 2 \sigma^2(r)}\right]\right\},
\label{eq:L2}
\end{eqnarray}
where $v_{\rm c}(r)$ is the circular speed at radius $r$, ln $\Lambda(r)$ is
the Coulomb logarithm, which we assume to be 4.0, $M_{\rm GC}$ is the mass of
one globular cluster. $\rho(r)$ is the density of the dark matter halo at
radius $r$ according to equation (\ref{eqnfw}) and $\sigma (r)$ is the
one-dimensional velocity dispersion of the halo. Solving this equation
numerically gives $t_{\rm fric} = 0.4$\,Gyr, which is very similar to the value
we directly get from the simulation.

\begin{figure}
\begin{center}
\epsfxsize=8.4cm
\epsfysize=7.0cm
\epsffile{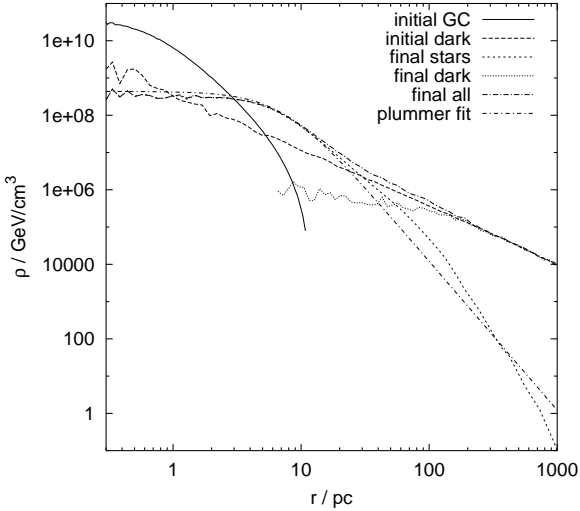}
\end{center}
\caption{The different density profiles of the UCD and the corresponding dark
matter distributions, after the ten globular clusters merged, but before and
after it has been put into an external potential. On top of the density profile
of the final stellar distribution we show the best fit plummer model.}
\label{figdenpro}
\end{figure}

The central compact object which results from the coalescence of the globular
clusters has a three-dimensional density profile which is shown in Fig.
\ref{figdenpro}. The profile is very similar to ones derived from observations
\cite{propris} as well as previous simulations (cf. Fig. 5 in Fellhauer \&
Kroupa (2002)). The density profile of the baryonic matter only (i.e. just the
material, which has been in the original globular clusters) can be well-fitted
with a Plummer model \cite{plummer}:
\begin{equation}
\rho(r)=\frac{3 M}{4 \pi b^3}\left(1+\frac{r^2}{b^2}\right)^{-5/2}.
\end{equation}

\begin{figure}
\begin{center}
\epsfxsize=8.4cm
\epsfysize=7.0cm
\epsffile{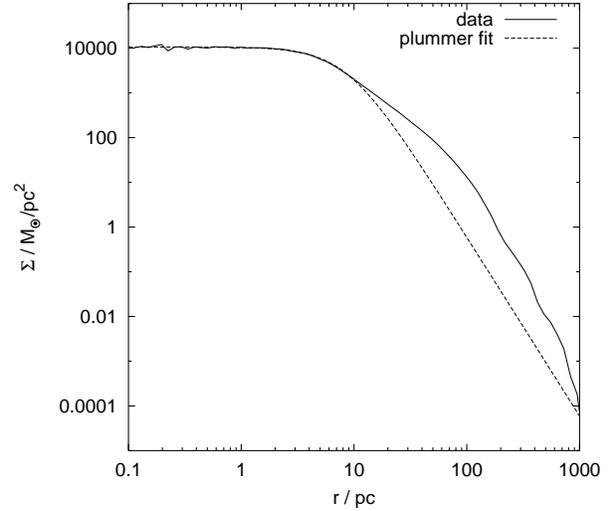}
\end{center}
\caption{Projected surface density profile of the UCD, after the ten globular
clusters merged, but before it has been put into an external potential. The
profile of the luminous matter together with the best fit plummer model is
plotted.}
\label{figsurpro}
\end{figure}

In our case $M = 1.45 \times 10^6$\,M$_{\odot}$ and $b = 7.26$\,pc. The
resulting object has a half mass radius of 25 pc and a central velocity
dispersion of 20\,km/s. The projected surface density profile can be seen in
Fig. \ref{figsurpro}. The fit is good in the inner parts and reasonable in the
outer parts. The equation for the fit has been found by Evans \& An (2005):
\begin{equation}
\Sigma(r) = \frac{A}{\pi\left(1 + r^2 / c^2\right)^2}
\end{equation}

Our fit gives $A=3.3\times10^4$\,M$_\odot/$pc$^2$ and $c=8.7$\,pc. It looks
fairly similar to observations and previous simulations [cf. Fig. 5 in
Ha\c{s}egan et al. (2005) or Fig. 6 in Fellhauer \& Kroupa (2002)]. If we
assume a mass to light ratio of three, we expect an absolute visual magnitude
of -11 for the final object. For comparison, a typical UCDs found in Fornax has
a central velocity dispersion of 22\,km/s, an absolute magnitude of -12 and
a half mass radius of 25\,pc \cite{ACS,drinkwater}. In order to form this
system by merging globular clusters one would need to consider many more
globulars to reach the observed luminosity. This turns out to be a crucial
point of this formation scenario because the presence of more than ten
globulars in such a halo is extremely rare \cite{sharina}.

\begin{figure}
\begin{center}
\epsfxsize=8.4cm
\epsfysize=7.0cm
\epsffile{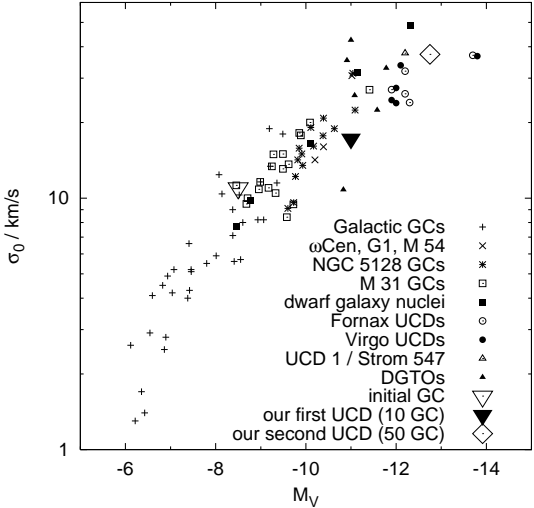}
\end{center}
\caption{Central velocity dispersion versus absolute visual magnitude for
various stellar systems. Observations are plotted together with the results
from our simulations. Data taken from Ha\c{s}egan et al. (2005); Evstigneeva et
al. (2007) and references therein.}
\label{figsm}
\end{figure}

In our second simulation, the fifty globular clusters merged completely within
a few million years. The resulting object had a central velocity dispersion of
37.3\,km/s and an absolute magnitude of -12.75, assuming a mass to light ratio
of three. Its position is marked in Fig. \ref{figsm} and roughly matches a UCD.

Interestingly, all dark matter particles are expelled from the core of the
final object and the inner initial cusp has been turned into a nearly constant
density core. Indeed, in our ten globular cluster simulation not even a single
dark matter particle is found within the inner 5\,pc. This effect has been
discussed in more detail in Read et al. (2006) and constitutes the second
reason for excluding the merging of globular clusters as a possible formation
mechanism for UCDs. In other words, the observed M/L ratio cannot possibly be
reached without dark matter. Only if the initial globular clusters contained
cuspy dark matter distributions would such a scenario stand a chance of
working, however current observations do not support this idea in the galactic
globular clusters.

Keeping in mind the observational uncertainty of the M/L ratios of UCDs, the
mechanism proposed by Fellhauer \& Kroupa (2006) to enhance the M/L ratio of
UCDs as well as resolution limits and other shortcomings of properly
determining the M/L ratio of our simulations in the next section, the reader
should note that the major issue and stronger argument against this formation
scenario is the total central luminosity.

\section{Are UCDs stripped disk galaxies?}
The second scenario we consider for the formation of UCDs is that they are the
remnants of stripped disk galaxies. For the numerical experiments performed in
this section, we build galaxy models using the following procedure. First, we
set up a spherical equilibrium NFW halo with structural parameters consistent
with predictions of the standard $\Lambda$CDM model \cite{stelios}. We include
a gaseous component of mass equal to a fraction $f_{\rm b}$ of the total halo
mass. The gas has originally the same radial distribution and a temperature
profile such that it is in hydrostatic equilibrium for an adiabatic equation
of state (EOS).

The gas component has a specific angular momentum distribution and spin
parameter consistent with values found for dark matter haloes within
cosmological N-body simulations \cite{bullock,andrea}. We constructed a dark
plus gaseous halo model with parameters that are expected to produce disks
similar to the Local Group galaxy M33 which has a small dense stellar nucleus.
The model parameters were: $M_{\rm vir} = 5 \times 10^{11}$ M$_{\odot}$,
$r_{\rm vir} = 167$ kpc, $v_{\rm vir} = 115$ km/s, $c=6.2$, baryonic fraction
$f_{\rm b}=6\%$ and spin parameter $\lambda = 0.1$.

The value of the concentration may seem a little bit low, when comparing it to
the results of Macci{\`o} et al. (2007) but it is the best fit for the M33
galaxy \cite{corbelli} which is the possible prototype progenitor for UCDs, we
want to test here. The value of the spin parameter is what one gets from
cosmological dark matter simulations. This net spin results from adding many
large, almost randomly oriented angular momenta of individual dark matter
particles. Uniformly rotating gas halos are normally not observed in
cosmological simulations, gas is accreted pretty much like the dark matter,
leading to random, large bulk motions and turbulence \cite{wise}. So one has to
note here that this setup is designed to produce cored disks for the subsequent
experiments and not necessarily an attempt to follow disk and core formation
realistically.

The hot gaseous halo is resolved with $2 \times 10^6$ particles of equal mass
$\sim 2 \times 10^4$ M$_\odot$. We sampled the dark matter halo with $2.2
\times 10^6$  particles having variable masses with the resolution increasing
towards the centre of the system \cite{zemp}. With a single-particle model one
would need about ten million particles to reach a comparable resolution in the
central region. This allows us to simulate the central dark matter cusp with
softening of 100 pc (this is the same for all dark matter and gas particles)
and particle mass of $\sim 4.4 \times 10^4$ M$_\odot$. The detailed description
of the initial conditions and results of the evolution of the disk are
presented in Kaufmann et al. (2006) and Kaufmann et al. (2007).

We use the parallel Tree+SPH code \textsc{Gasoline} \cite{Wadsley}, which is an
extension of the pure N-Body gravity code \textsc{Pkdgrav} developed by Stadel
(2001). It uses a 32 particles smoothing kernel and an artificial viscosity
which is the shear reduced version \cite{Bals} of the standard Monaghan (1992)
artificial viscosity. \textsc{Gasoline} uses a spline kernel with compact
support for the softening of the gravitational and SPH quantities. The energy
equation is solved using the asymmetric formulation, which is shown to yield
very similar results compared to the entropy conserving formulation but
conserves energy better \cite{Wadsley}. The code includes radiative cooling for
a primordial mixture of helium and (atomic) hydrogen. Because of the lack of
molecular cooling and metals, the efficiency of our cooling functions drops
rapidly below 10\,000 K, but we adopt a temperature floor of $T_{\rm f} =$
15\,000 K, to crudely mimic the effect of heating sources such as supernovae
explosions and radiation (e.g. ultraviolet) backgrounds. These simulations were
performed without star formation. We do not expect them to suffer from the SPH
problems pointed out by Agertz et al. (2007) because for the disk forming part
of the simulations we are adopting a large temperature floor and do not aim to
simulate the actual multiphase ISM, where the problems would occur. The gas
cools slowly to form a disk and stars are formed at a rather low threshold,
therefore there is problem with our simulations (Oscar Agertz, private
communication). For a detailed discussion on the performance of SPH for the
stripping part of the simulations we defer the interested reader to McCarthy
(2007). 

\begin{figure}
\begin{center}
\epsfxsize=8.4cm
\epsfysize=7.0cm
\epsffile{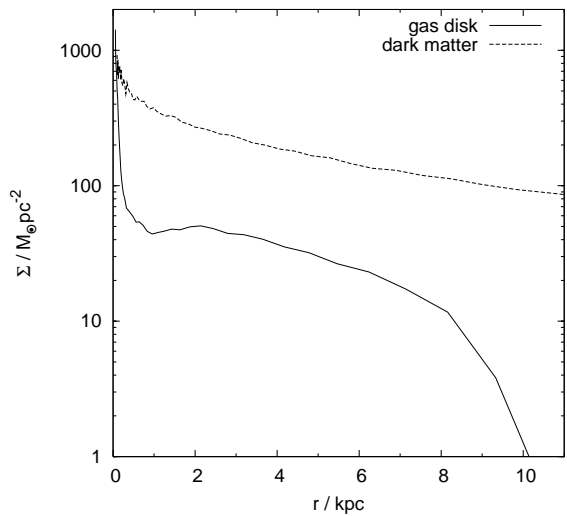}
\end{center}
\caption{The logarithmic surface density of the M33 gas disk is plotted after
2.5 Gyr. The nucleus in the centre is clearly visible. The dark matter
distribution is plotted on top for comparison.}
\label{nucleus}
\end{figure}

Due to cooling the hot gas halo loses its hydrostatic equilibrium quite
quickly. An inner gaseous disk rapidly forms out of cooling gas coming from a
nearly spherical region close to the halo centre (within $\sim 10$ kpc). After
2.5 Gyr of evolution  the disk attains a near exponential surface density
profile over a large fraction of its extent, except within a few hundred
parsecs from the centre, where gas inflows along an inner bar produce
a dense nucleus which can be seen in Fig. \ref{nucleus}.

The nucleus has a mass of $\sim 3 \times 10^{7}$ M$_{\odot}$, which is $\sim
0.6 \%$ of the total disk. Central dense stellar nuclei have been seen in
several late-type spiral galaxies, for example, in M33 \cite{regan}, but also
in other late-type spirals \cite{Carollo1,Carollo2}. Recent observations by
Ferrarese et al. (2006) show that they are observed in 50 - 80\% of low- and
intermediate-luminosity galaxies.

We put the galaxy into the same artificial NFW potential resembling the Virgo
cluster that we used before. We start with the disk at pericentre, 100 kpc far
from the centre of the static potential, and give it a perpendicular velocity
of 260 km/s, such that it should reach an apocentre at 30 kpc. The plane of the
disk is tilted by 45$^\circ$ with respect to the plane of the orbit. On this
orbit the outer galactic halo and gas are quickly stripped from the nucleus.

\begin{figure}
\begin{center}
\epsfxsize=8.4cm
\epsfysize=7.0cm
\epsffile{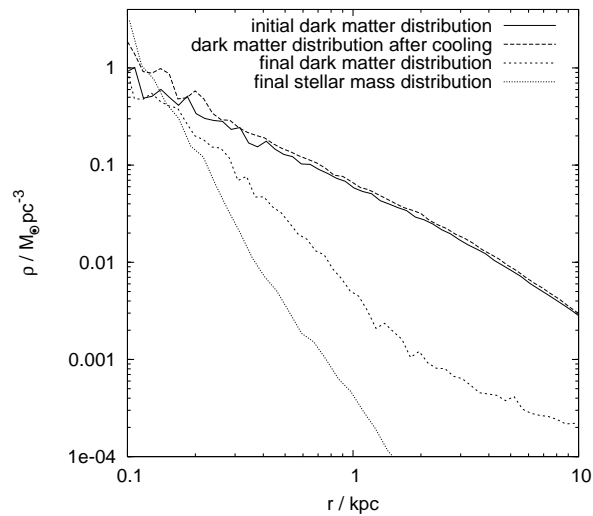}
\end{center}
\caption{The dark matter distributions of the disk model in various stages of
its evolution, together with the final baryonic (stellar) mass 
distribution.}
\label{dmden}
\end{figure}

The resolution of our hydrodynamic simulation is not sufficient to allow a
meaningful comparison with the observations. However the half light radius,
luminosity and dark matter content all agree well with observations up to the
given resolution limits (see Fig. \ref{dmden}), the mass to light ratio within
the optical radius ($\sim$ 50pc) is about 4. and the UCD is dark matter
dominated up to the resolution limit.

\subsection{Defining the critical disruption zone and the affects of baryons
  on survival versus disruption}

If the orbit of the galaxy does not penetrate the centre of the cluster then
the disk would not be stripped and the object would not be classified as a UCD.
If the progenitors are nucleated disks which are tidally transformed into
nucleated spheroidals dE(N) via galaxy harassment \cite{moore2} then we might
expect a transition region from central UCDs to outer dE(N). Since UCDs are all
located close to the centres of clusters and groups this allows us to explore
the conditions under which complete disruption occurs and how that is affected
by the amount of dissipation (steepening of the central potential).

In what follows, we performed an extensive set of simulations to identify the
orbits on which the disk galaxy is completely stripped and on which it remains
intact. The first case would correspond to a UCD and the second to a dE(N).
The central potential of the SPH galaxy is deepened significantly due to
dissipation. The object will be harder to completely disrupt as compared to the
initial dark matter halo of the model for example. We therefore make a
systematic study of the orbits that lead to disruption versus survival for the
nucleated M33 galaxy model and for the initial uncompressed dark matter halo -
the latter simulations define the optimum scenario for complete disruption.

These simulations all start at their respective apocentres and run for the same
physical time (5 Gyr) in the cluster potential, and explore different orbital
apocentres and pericentres. If stripping of disk galaxies is the correct
mechanism for forming UCDs and dE(N), there must be a very sudden transition
between these two cases. This is because we do not observe an intermediate
object with features lying in between these very distinct two types of
galaxies.

\begin{figure}
\begin{center}
\epsfxsize=8.4cm
\epsfysize=7.0cm
\epsffile{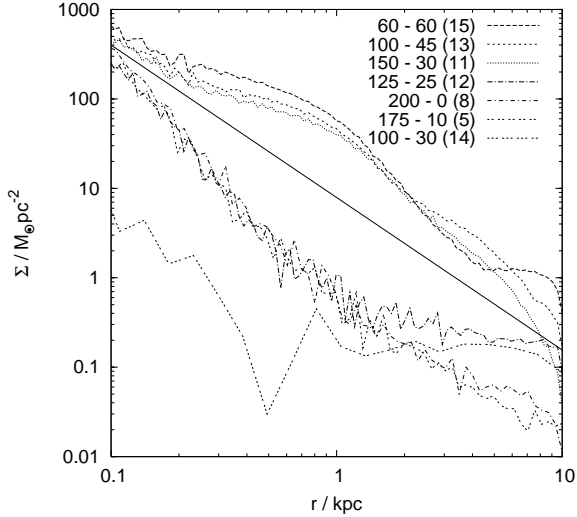}
\end{center}
\caption{Surface brightness profiles of the stripped disk after orbiting an
artificial Virgo cluster potential for 5 Gyr. The numbers next to the hyphen in
the legend give the apo- and pericentres of the various orbits in kpc and the
numbers within the brackets give the number of orbits which were completed
during the 5 Gyrs. The solid line indicates the dividing line between survival
and disruption.}
\label{striporbit}
\end{figure}

Fig. \ref{striporbit} presents surface brightness profiles of the disk galaxy
after orbiting the static cluster potential for 5 Gyr on various orbits. A
sudden transition is clearly seen in this plot: almost two orders of magnitude
in surface brightness lie between the (150 - 30) and the (125 - 25) orbit at
the radii of interest, so the disk either survives or is completely disrupted.

\begin{figure}
\begin{center}
\epsfxsize=8.4cm
\epsfysize=7.0cm
\epsffile{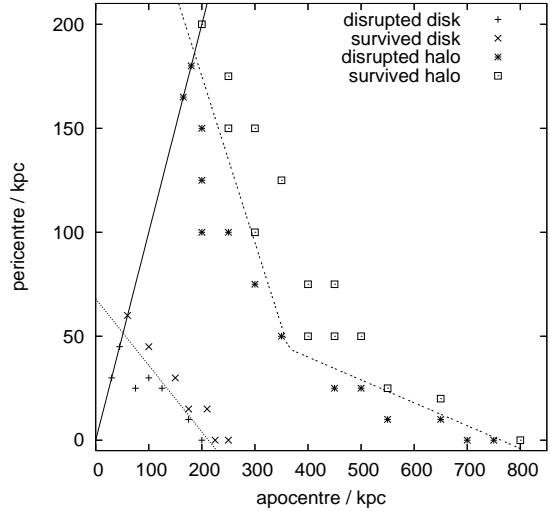}
\end{center}
\caption{Disrupted and non disrupted gas disks and dark matter halo nuclei as a
function of apo- and pericentre for all the disk and the dark matter only
simulations we have done. The solid line is the bisector and the two dotted
ones are the dividing lines between disruption and survival.}
\label{allsims}
\end{figure}

Fig. \ref{allsims} presents a scatter plot of apocentres versus pericentres for
all cases we simulated along with indications about the fate of the orbiting
system. The dividing line between disrupted and surviving disks is given by
\begin{equation}
r_{\rm peri} = A r_{\rm apo} + B
\end{equation}
with A = -0.32 and B = 68.

We also repeated the above stripping experiments with the dark matter halo only
without the gas disk. Results of these experiments are included in Fig.
\ref{allsims}.
In this case the transition is not as sudden as in the disk case therefore we
define a halo to be completely disrupted, if less than 0.1 \% of the virial
mass is still bound. Defining complete disruption by the bound mass getting
below a certain threshold is the most objective way to do so. The value of
0.1\%, may seem arbitrary, but it is actually the value, which gave the most
convincing values for the given resolution and algorithm. To estimate the mass
which is still bound we used the group finding algorithm \textsc{Skid}
\cite{stadel}. There is a clear separation between survival and disruption that
depends only on the apocentre and pericenter. The theoretical explanation for
the orbits that lead to complete disruption is fairly intricate and will be
discussed in a later paper.

The dividing line between disrupted and surviving haloes is given roughly by
two straight lines:
\begin{eqnarray}
r_{\rm peri} & = & A r_{\rm apo} + B \nonumber \\
r_{\rm peri} & = & C r_{\rm apo} + D
\end{eqnarray}
with A = -0.11, B = 84.0, C = -0.8, D = 335.0, with the transition between them
at $r = (D - B) / (A - C) \sim$ 370 kpc.

Given these empirically determined survival/disruption curves, we can sample
orbits from a cosmological cluster mass CDM halo and determine the expected
radial profile of completely disrupted haloes (UCDs) in each of these two
cases. From the cosmological simulation we chose the one cluster which was not
perturbed by neighbouring clusters, had only a single nucleus and came closest 
in $M_{\rm vir}$ and $r_{\rm vir}$ to the Virgo cluster. 

\begin{figure}
\begin{center}
\epsfxsize=8.4cm
\epsfysize=7.0cm
\epsffile{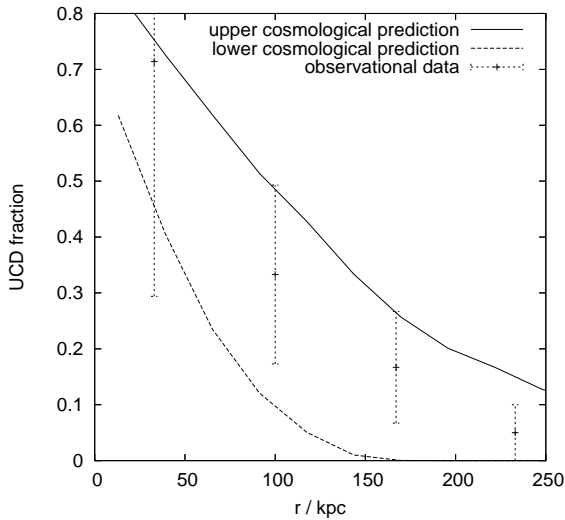}
\end{center}
\caption{The ratio of the number of UCDs to the number of progenitors, which is
sum of the number of UCDs and the number of dE(N)s, as a function of radius.
The upper cosmological prediction comes from the disruption simulations using 
the pure dark matter subhalo whilst the lower cosmological prediction comes
from the SPH simulations with the dissipated gas disk. Observational data is
taken from Jones et al. (2006).}
\label{fraction}
\end{figure}

We evolved the halo for 5 Gyr and binned all particles according to their
projected distance from the centre at the last simulation output. For each bin
we determined the fraction of particles whose orbits during the 5 Gyr of
evolution lies below the respective dividing line in Fig. \ref{allsims}. The
orbits have been determined in a way that the pericentre is the closest point
the particle gets to on all snapshot and the apocentre is the furthest point.
In Fig. \ref{fraction} this fraction is then compared to its observed
counterpart in nature. This is the fraction of the number of UCDs to the number
of progenitors, which is the sum of the number of UCDs and the number of
dE(N)s, so:
\begin{equation}
f_{\rm UCD} = {n_{\rm UCD} \over n_{\rm progenitor}} = {n_{\rm UCD} \over
n_{\rm UCD} + n_{\rm dE(N)s}}
\end{equation}

Results from the pure dark matter stripping experiments correspond to the upper
cosmological prediction (The final radial distribution of UCDs is more extended
since the dark matter halo is not compressed due to dissipation and disrupts at
larger clustercentric radii). The SPH disk galaxy stripping simulations
correspond to the lower cosmological prediction. We have to remind the reader
again that our prototype progenitor was M33 that has a concentration, which
might look a bit low according to Macci{\`o} et al. (2007). Therefore the
subhalo disruption and the UCD fractions might be slightly overestimated.
However, one could image the progenitor being heavier, therefore less
concentrated and thus showing the same disruption behaviour as the haloes we
present. Unfortunately we did not have the computational resources to explore
parameter space any further. The observational data have been obtained in the
following way. The projected distances of UCDs and dE(N)s in arcsec (Jones et
al. 2006) are converted into kpc assuming that the Virgo and Fornax cluster lie
at distances of 16 and 20 Mpc, respectively. In a $\Lambda$CDM universe, haloes
are approximately self similar so we can compare results by scaling everything
with respect to the virial radius. After rescaling, we binned the occurrence of
the UCDs and dE(N)s according to their rescaled projected distance to their
mutual galaxy cluster centre into common bins. The content of each bin is
associated with a Poissonian error. Finally, we divide the number of UCDs in
each bin by the sum of the numbers of UCDs and dE(N)s and propagate the errors
accordingly. From comparing our two cosmological prediction with the data, we
conclude that (a) both UCDs and dENs originate from the same progenitors and
(b) this progenitor must have properties in between the disk galaxy and the
pure dark matter halo we adopted in our simulations. One must bear in mind
though that we only have position data for 15 UCDs from the literature.
Nonetheless, and limitations notwithstanding, our cosmological predictions
agree well with the data but further observations will allow a better test of
this model.

\section{Conclusions}
UCDs have similar properties as massive globular clusters or the nuclei of
nucleated galaxies. Recent observations of a high dark matter content and their
steep spatial distribution within groups and clusters give us new clues as to
their origins. We perform $N$-body simulations and compare two possible
mechanisms for their formation: the merging of globular clusters in the centre
of a dark matter halo, or the massively stripped remnant of a nucleated galaxy.
Our simulations reveal, that a swarm of ten as well as one of fifty globular
clusters born in a Fornax dwarf spheroidal size dark matter halo will normally
evolve to an object, which has the same density profile as an UCD. We performed
a second series of simulations in which a disk galaxy experiences mass loss
processes inside a cluster environment. The disk is entirely stripped and the
remaining nucleus exhibits all of the observed UCD properties. Both models
produce density profiles as well as the half light radii that can fit the
observational constraints very well.

However, we show that the first scenario produces UCDs that are underluminous
and contain no dark matter - the sinking process ejects most of the dark matter
particles from the halo centre. The stars, which consist the globulars, would
replace the dark matter particles and expel them from the centre of the halo
\cite{amr,merritt} lowering the M/L ratio beyond observational constraints
\cite{ACS}. The other drawbacks of this model is that it is very unlikely to
have so many globular cluster in a halo of that size. It becomes apparent in
Fig. \ref{figsm} that it is not sufficient to merge only ten globulars:
approximately fifty are necessary. However such a halo may not exist
\cite{sharina}.

Stripped nuclei give a more promising explanation, especially if the nuclei
form via the sinking of gas, funneled down inner galactic bars, since this
process enhances the central dark matter content. Even when the entire disk is
stripped away, the nucleus remains intact and can be dark matter dominated. The
total disruption of the galaxy beyond the nuclei only occurs on certain orbits
and depends on the amount of dissipation during nuclei formation. By comparing
the total disruption of CDM subhaloes in a cluster potential we show that this
model also leads to the observed spatial distribution of UCDs. This scenario
can be tested in more detail with larger data sets.

\section*{Acknowledgements}
It is a pleasure to thank Monica Ha\c{s}egan, Bryn Jones, Katya Evstigneeva and
Michael Drinkwater for their kindness to send us data to help with our figures.
SK is supported by a Kavli Institute for Particle Astrophysics and Cosmology
Postdoctoral Fellowship at Stanford University. All computations were made on
the zBox and zBox2 supercomputers at the University of Z\"urich. Special 
thanks to Doug Potter for bringing zBox2 to life. Tobias Goerdt is a Golda Meir
fellow.

\label{lastpage}


\begin{thebibliography}{}

\bibitem[\protect\citename{Agertz et al. }2007]{Oscar}
Agertz O, Moore B, Stadel J. et al, 2007, MNRAS, 380, 963

\bibitem[\protect\citename{Balsara }1995]{Bals}
Balsara D. S, 1995, Journal of Computational Physics, 121, 357

\bibitem[\protect\citename{Bekki et al. }2001]{Bekki1}
Bekki K, Couch W. J, Drinkwater M. J, 2001, ApJ, 552, 105

\bibitem[\protect\citename{Bekki et al. }2003]{Bekki2}
Bekki K, Couch W. J, Drinkwater M. J, 2003, MNRAS, 344, 399

\bibitem[\protect\citename{B\"oker et al. }2002]{boeker1}
B\"oker T, Laine S, van der Marel R. P. et al, 2002, AJ, 123, 1389

\bibitem[\protect\citename{B\"oker et al. }2004]{boeker2}
B\"oker T, Sarzi M, McLaughlin D. E. et al, 2004, AJ, 127, 105

\bibitem[\protect\citename{Bullock et al. }2001]{bullock}
Bullock J. S, Dekel A, Kolatt T. S. et al, 2001, ApJ, 555, 240

\bibitem[\protect\citename{Carollo }1999]{Carollo1}
Carollo C. M, 1999, ApJ, 523, 566

\bibitem[\protect\citename{Carollo, Stiavelli \& Mack }1998]{Carollo2}
Carollo C. M, Stiavelli M, Mack J, 1998, AJ, 116, 68

\bibitem[\protect\citename{Chandrasekhar }1943]{chandrasekhar}
Chandrasekhar S, 1943, ApJ, 97, 255

\bibitem[\protect\citename{Corbelli }2003]{corbelli}
Corbelli E, 2003, MNRAS, 342, 199

\bibitem[\protect\citename{De Propris et al. }2005]{propris}
De Propris R, Phillipps S, Drinkwater M. J, 2005, ApJ, 623, 105

\bibitem[\protect\citename{Drinkwater et al. }2000]{drinkwater2}
Drinkwater M. J, Jones J. B, Gregg M. D, Phillipps S, 2000, Publ. Astron. Soc.
Australia, 17, 227

\bibitem[\protect\citename{Drinkwater et al. }2003]{drinkwater}
Drinkwater M. J, Gregg M. D, Hilker M et al, 2003, Nature, 423, 519

\bibitem[\protect\citename{El-Zant, Shlosman \& Hoffman }2001]{amr}
El-Zant A, Shlossman I, Hoffman Y, 2001, ApJ 560, 636

\bibitem[\protect\citename{Evans \& An }2005]{evans}
Evans N. W, An J, 2005, MNRAS, 360, 492

\bibitem[\protect\citename{Evstigneeva et al. }2007]{gregg}
Evstigneeva E. A, Gregg M. D, Drinkwater M. J, Hilker M, 2007, AJ, 133, 1722

\bibitem[\protect\citename{Fellhauer \& Kroupa }2002]{fellhauer}
Fellhauer M, Kroupa P, 2002, MNRAS, 330, 642

\bibitem[\protect\citename{Fellhauer \& Kroupa }2006]{fellhauer2}
Fellhauer M, Kroupa P, 2006, MNRAS, 367, 1577

\bibitem[\protect\citename{Ferrarese }2006]{ferrarese}
Ferrarese L, Cote P, Bonta E. D. et al, 2006, ApJ, 644, 21

\bibitem[\protect\citename{Goerdt et al. }2006]{goerdt}
Goerdt T, Moore B, Read J. I, Stadel J, Zemp M, MNRAS, 368, 1073

\bibitem[\protect\citename{Ha\c{s}egan et al. }2005]{ACS}
Ha\c{s}egan M, Jordan A, Cote P. et al, 2005, ApJ, 627, 203

\bibitem[\protect\citename{Hernquist }1990]{abc}
Hernquist L, 1990, ApJ, 356, 359

\bibitem[\protect\citename{Hilker et al. }1999]{hilker}
Hilker M, Infante L, Viera G. et al. 1999, A\&AS, 134, 75

\bibitem[\protect\citename{Jones }2006]{jones}
Jones J. B, Drinkwater M. J, Jurek R. et al, 2006, AJ, 131, 312

\bibitem[\protect\citename{Kazantzidis, Magorrian \& Moore }2004]{stelios}
Kazantzidis S, Magorrian J, Moore B, 2004, ApJ, 601, 37

\bibitem[\protect\citename{Kazantzidis, Moore \& Mayer }2003]{steliosucd}
Kazantzidis S, Moore B, Mayer L, 2003, astro-ph/0307362

\bibitem[\protect\citename{Kaufmann et al. }2006]{Kaufmann2006a}
Kaufmann T, Mayer L, Wadsley J, Stadel J, Moore B, 2006, MNRAS, 370, 1612

\bibitem[\protect\citename{Kaufmann et al. }2007]{Kaufmann2006b}
Kaufmann T, Mayer L, Wadsley J, Stadel J, Moore B, 2007, MNRAS, 375, 53

\bibitem[\protect\citename{King }1966]{king}
King I. R, 1966, AJ, 71, 64

\bibitem[\protect\citename{Macci{\`o} et al. }2007]{andrea}
Macci{\`o} A. V, Dutton A. A, van den Bosch F. C, Moore B, Potter D, Stadel J,
2007, MNRAS, 378, 55

\bibitem[\protect\citename{Mateo }1998]{Mateo98}
Mateo M. L, 1998, ARAA, 36, 435

\bibitem[\protect\citename{Merritt et al. }2004]{merritt}
Merritt D, Piatek S, Zwart S. P, Hemsendorf M, 2004, ApJ, 608, 25

\bibitem[\protect\citename{McCarthy et al. }2008]{mccarthy}
McCarthy I. G, Frenk C. S, Font A. S. et al, 2008, MNRAS, 383, 593

\bibitem[\protect\citename{Michie }1963]{michie}
Michie R. W, 1963, MNRAS, 125, 127

\bibitem[\protect\citename{Michie \& Bodenheimer }1963]{bodenheimer}
Michie R. W, Bodenheimer P. H, 1963, MNRAS, 126, 269

\bibitem[\protect\citename{Moore et al. }1996]{moore2}
Moore B, Katz N, Lake G, Dressler A, Oemler A, 1996, Nature, 379, 613
 
\bibitem[\protect\citename{Moore et al. }1999]{moore}
Moore B, Quinn T, Governato F, Stadel J, Lake G, 1999, MNRAS, 310, 1147

\bibitem[\protect\citename{Monaghan }1992]{Monagh}
Monaghan J. J, 1992, ARAA, 30, 543

\bibitem[\protect\citename{Navarro, Frenk \& White }1996]{nfw}
Navarro J. F, Frenk C. S, White S. D. M, 1996, ApJ, 462, 563

\bibitem[\protect\citename{Oh \& Lin }2000]{oh}
Oh K. S, Lin D. N. C, 2000, ApJ, 543, 620

\bibitem[\protect\citename{Oh, Lin \& Aarseth }1995]{oh2}
Oh K. S, Lin D. N. C, Aarseth S. J, 1995, ApJ, 422, 142

\bibitem[\protect\citename{Oh, Lin \& Richer }2000]{oh3}
Oh K. S, Lin D. N. C, Richer H. B, 2000, ApJ, 531, 727

\bibitem[\protect\citename{Phillipps et al. }2001]{phillipps}
Phillipps S, Drinkwater M. J, Gregg M. D, Jones J. B, 2001, ApJ, 560, 201

\bibitem[\protect\citename{Plummer }1911]{plummer}
Plummer H. C, 1911, MNRAS, 71, 460

\bibitem[\protect\citename{Read et al. }2006]{read}
Read J. I, Goerdt T, Moore B. et al, 2006, MNRAS, 373, 1451

\bibitem[\protect\citename{Regan \& Vogel }1994]{regan}
Regan M. W, Vogel S. N, 1994, ApJ, 434, 536

\bibitem[\protect\citename{Sharina, Puzia \& Makarov }1996]{sharina}
Sharina M. E, Puzia T. H, Makarov D. I, 2005, A\&A, 442, 85

\bibitem[\protect\citename{Stadel }2001]{stadel}
Stadel J, 2001, PhD thesis, Univ. Washington

\bibitem[\protect\citename{Wadsley, Stadel \& Quinn }2004]{Wadsley}
Wadsley J, Stadel J, Quinn T, 2004, NewA, 9, 137

\bibitem[\protect\citename{Walcher et al. }2005]{walcher}
Walcher C. J, van der Marel R. P, McLaughlin D. et al, 2005, ApJ, 618,
237, erratum-ibid, 2005, 635, 741

\bibitem[\protect\citename{Wise \& Abel }2007]{wise}
Wise J. H, Abel T, 2007, ApJ, 665, 899

\bibitem[\protect\citename{Zemp et al. }2007]{zemp}
Zemp M, Moore B, Stadel J, Carollo C. M, Madau P, 2007, arXiv:0710.3189
[astro-ph]

\end{thebibliography}
\end{document}